\documentclass[aps,prl,twocolumn,showpacs,superscriptaddress]{revtex4-1}
\pdfoutput=1
\usepackage{amsfonts} 
\usepackage{amsmath}
\usepackage{amssymb}
\usepackage{bm}
\usepackage[usenames, dvipsnames]{color} 
\usepackage{dcolumn}
\usepackage{epic} 
\usepackage{epsfig}
\usepackage{graphicx}
\usepackage[abs]{overpic}
\usepackage[lofdepth,lotdepth,caption=false]{subfig}

\usepackage[breaklinks,colorlinks = true,linkcolor = blue,urlcolor=blue,citecolor=blue]{hyperref}
\usepackage{soul}
\usepackage{times}
\usepackage{ulem}
\usepackage{wrapfig} 
\usepackage{xy} 

\newcommand{\beq}{\begin{equation}}
\newcommand{\eeq}{\end{equation}}
\newcommand{\beqa}{\begin{eqnarray}}
\newcommand{\eeqa}{\end{eqnarray}}

\newcommand{\hhc}{hyperhoneycomb}
\newcommand{\hn}[1]{$\mathcal{H}\text{--}#1$}
\newcommand{\bk}{\mathbf{k}}

\newcommand{\Rmnum}[1]{\expandafter\@slowromancap\romannumeral #1@}

\begin{document}

\title{Topological spinon semimetals and gapless boundary states in three dimensions}


\author{Robert Schaffer}
\author{Eric Kin-Ho Lee}
\affiliation{Department of Physics and Center for Quantum Materials,
University of Toronto, Toronto, Ontario M5S 1A7, Canada.}
\author{Yuan-Ming Lu}
\affiliation{Department of Physics, University of California,
Berkeley, CA 94720} \affiliation{Materials Sciences Division,
Lawrence Berkeley National Laboratory, Berkeley, CA 94720}
\author{Yong Baek Kim}
\affiliation{Department of Physics and Center for Quantum Materials,
University of Toronto, Toronto, Ontario M5S 1A7, Canada.}
\affiliation{School of Physics, Korea Institute for Advanced Study, Seoul 130-722, Korea.}

\begin{abstract}
  Recently there has been much effort in understanding topological
  phases of matter with gapless bulk excitations, which are
  characterized by topological invariants and protected intrinsic
  boundary states. Here we show that topological semimetals of
  Majorana fermions arise in exactly solvable Kitaev spin models on a
  series of three dimensional lattices. The ground states of these
  models are quantum spin liquids with gapless nodal spectra of bulk
  Majorana fermion excitations. It is shown that these phases are
  topologically stable as long as certain discrete symmetries are
  protected. The corresponding topological indices and the gapless
  boundary states are explicitly computed to support these results. In
  contrast to previous studies of non-interacting systems, the phases
  discussed in this work are novel examples of gapless topological
  phases in interacting spin systems.
\end{abstract}
\date{\today}
\maketitle

\paragraph*{Introduction}
Theoretical prediction and experimental realization of topological
insulators\cite{Hasan2010,Hasan2011,Qi2011} (TIs) pushes our
understanding of topological phenomena in condensed matter physics to
a new level. Recently it was revealed that analogs of TIs exist in a
large class of interacting boson and spin systems, dubbed ``symmetry
protected topological phases''\cite{Chen2013,Senthil2014}. These topological phases are analogs of one another due to the existence of gapless surface states protected by symmetries, in spite of an energy gap for bulk
excitations. Similar to TIs, a rich topology also exists in semimetals
of weakly-interacting electrons featuring protected boundary
excitations\cite{Turner2013,Vafek2014}, such as Weyl semimetals with
surface Fermi arcs\cite{Wan2011}. This raises a natural question: are
there analogs of topological semimetals in interacting boson/spin
systems, which harbor both gapless bulk excitations and protected
surface states? Here we provide a positive answer to this question, in
the ground states of the Kitaev model\cite{Kitaev2006} on a series of
three-dimensional trivalent lattices.

Motivated by recent discovery of the \hhc{} (\hn{0}) and harmonic honeycomb (\hn{1})
iridates\cite{Takayama2014,Modic2014}, the Kitaev model on these
lattices have been examined\cite{Mandal2009,Lee2014,Kimchi2013,Nasu2014} and gapless $\mathbb{Z}_2$ spin-liquids
with one-dimensional spinon nodal rings were found to be plausible
ground states of these models\cite{Mandal2009,Lee2014,Kimchi2013}. In addition, a three-dimensional $\mathbb{Z}_2$ spin liquid with a two-dimensional Fermi surface was explored on the hyperoctagon lattice\cite{Hermanns2014}. In
this work, we show that the Majorana spinon nodal rings in the bulk
of the gapless spin liquids on the \hn{n} lattices are topologically stable.  Moreover, due
to the bulk-boundary correspondence\cite{Matsuura2013}, these spin
liquids exhibit protected gapless surface states in the form of
dispersionless zero-energy flat bands.

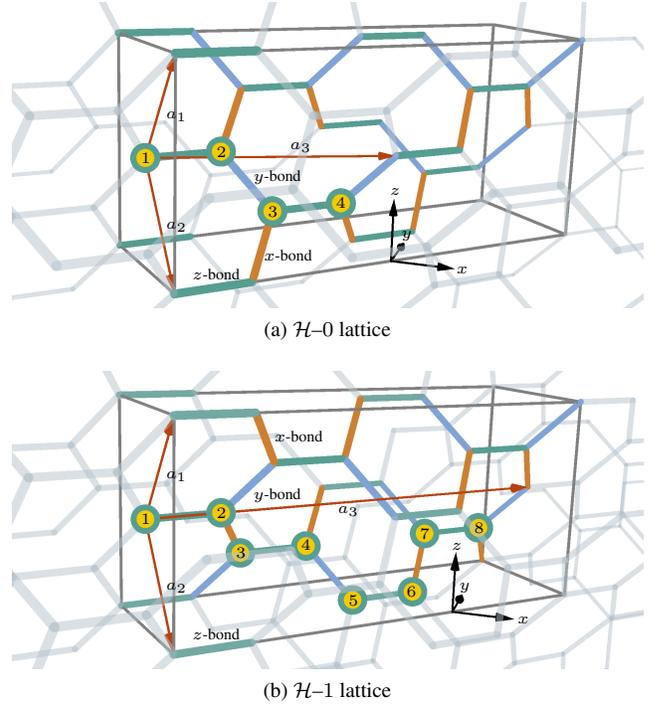
\begin{figure}[!htbp]
  \centering
  \setlength\fboxsep{0pt}
  \setlength\fboxrule{0pt}
  \subfloat[][\hn{0} lattice]{
    \label{fig:h0_lattice}
    \fbox{\begin{overpic}[width=3.3in,clip=true,grid=false,trim=110 565 185 630]{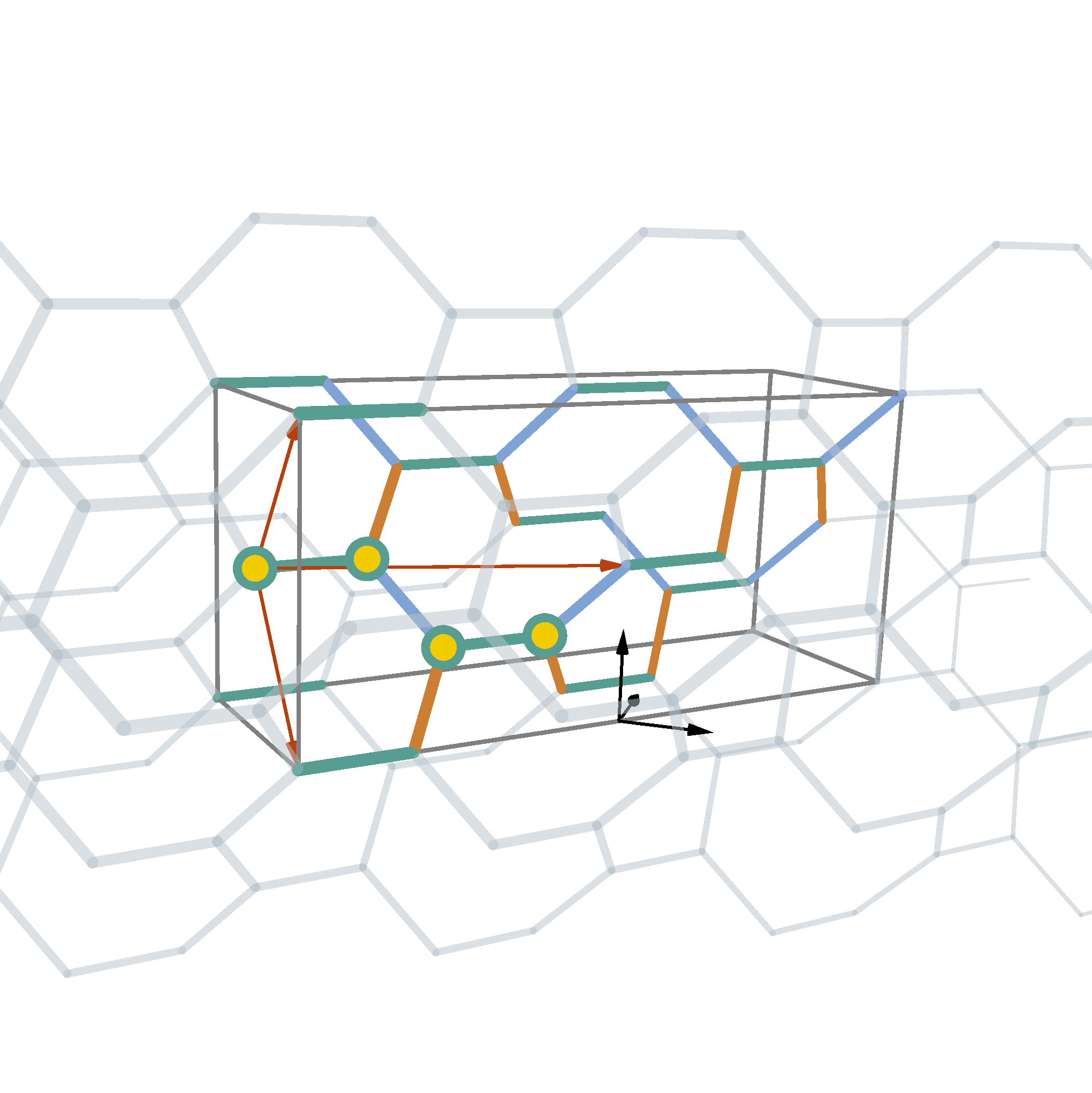}
        \put(48.5,54.9){$\scriptstyle{1}$}
        \put(77.1,57){$\scriptstyle{2}$}
        \put(96.6,34.5){$\scriptstyle{3}$}
        \put(122.3,37.8){$\scriptstyle{4}$}
        \put(142.7,43){$\scriptstyle{z}$}
        \put(147,26.5){$\scriptstyle{y}$}
        \put(168,13){$\scriptstyle{x}$}
        \put(58,72){$\scriptstyle{a_1}$}
        \put(58,30){$\scriptstyle{a_2}$}
        \put(105,60){$\scriptstyle{a_3}$}
        \put(68,11){$\scriptstyle{z\text{-bond}}$}
        \put(95,18){$\scriptstyle{x\text{-bond}}$}
        \put(91,48){$\scriptstyle{y\text{-bond}}$}
      \end{overpic}}
  }

  \subfloat[][\hn{1} lattice]{
    \label{fig:h1_lattice}
    \fbox{\begin{overpic}[width=3.3in,clip=true,grid=false,trim=60 550 200 650]{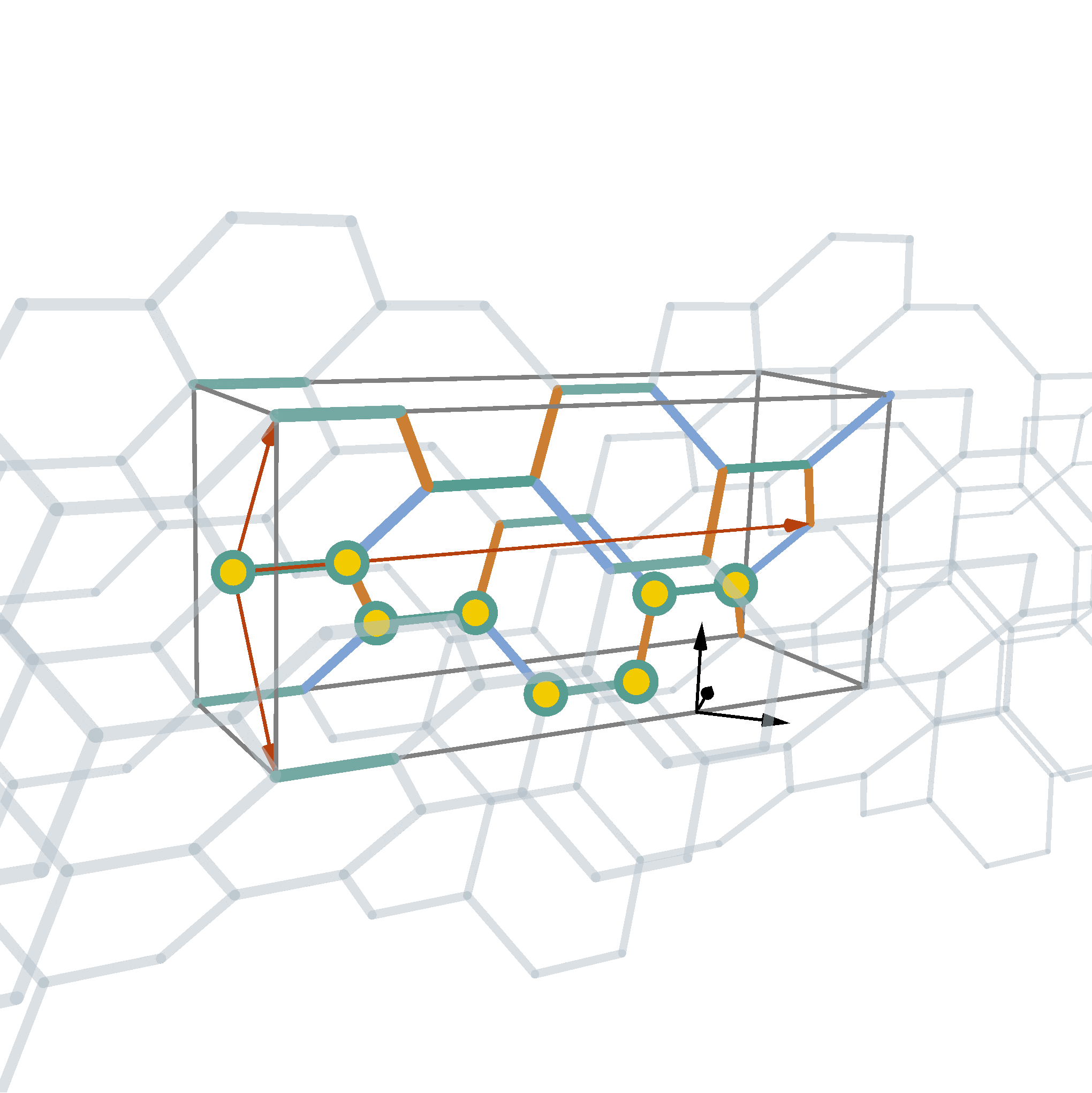}
        \put(48.5,54.9){$\scriptstyle{1}$}
        \put(77.1,57){$\scriptstyle{2}$}
        \put(84.5,42){$\scriptstyle{3}$}
        \put(109.1,44.8){$\scriptstyle{4}$}
        \put(127,24){$\scriptstyle{5}$}
        \put(149,27.3){$\scriptstyle{6}$}
        \put(154,49){$\scriptstyle{7}$}
        \put(174.6,51.5){$\scriptstyle{8}$}
        \put(166,45){$\scriptstyle{z}$}
        \put(169,30){$\scriptstyle{y}$}
        \put(191,17){$\scriptstyle{x}$}
        \put(58,72){$\scriptstyle{a_1}$}
        \put(58,30){$\scriptstyle{a_2}$}
        \put(123,59){$\scriptstyle{a_3}$}
        \put(68,11){$\scriptstyle{z\text{-bond}}$}
        \put(99,86){$\scriptstyle{x\text{-bond}}$}
        \put(91,64){$\scriptstyle{y\text{-bond}}$}
      \end{overpic}}
  }
  \caption{\label{fig:lattices}(Color online) Unit cell, lattice
    vectors, sublattices, and coordinate systems for the \hn{0} and
    \hn{1} lattices.  The conventional (orthorhombic) unit cells are
    drawn, while the sublattices in the primitive unit cells are
    labeled from $1$ to $4n+4$.  The $x$-, $y$-, and $z$-bonds within
    the conventional unit cell as defined in the Kitaev model are
    colored in orange, blue, and turquoise respectively.}
\end{figure}

\paragraph*{Solution to Kitaev model:} We first examine the bulk
properties of the ground states on these lattices. Given a particular
choice of the vectors $\hat{x}$, $\hat{y}$ and $\hat{z}$ (see
Fig. \ref{fig:lattices}), we define $x$, $y$ and $z$ bonds as those
which are perpendicular to the associated directions. With this
definition, each site shares one bond of each type with one of its
three neighbours. As such, we define the Kitaev Hamiltonian on these
lattices as
\begin{eqnarray}
  \mathbf{H} = \sum_{\langle ij \rangle \in \alpha} J_{\alpha} \mathbf{S}_i^{\alpha}
  \mathbf{S}_j^{\alpha},
\end{eqnarray}
where $\alpha$ denotes the bond type of bond $ij$, and the sum runs
over nearest neighbour bonds. We take $J_\alpha$ to be the same over
all bonds of type $\alpha$ and $J_x$ = $J_y$ for simplicity, which
preserves the crystal symmetries of the lattices.

This model can be solved exactly by introducing four Majorana spinons
$\{b^x,b^y,b^z,c\}$ at each site and replacing
$\mathbf{S}_i^\alpha=ib_i^\alpha c_i$\cite{Kitaev2006}. The operators
$u_{ij} = ib_i^\alpha b_j^\alpha = -u_{ji}$ (where $ij$ is an $\alpha$
bond) commute with one another and the Hamiltonian, thus they define
conserved quantities that take on the values of $\pm 1$ on each
bond. These $u_{ij}$ are not gauge invariant.  However, products of
these operators over closed loops, which correspond to fluxes of the
$Z_2$ gauge field, are gauge invariant\cite{Kitaev2006}. By choosing a
configuration of $\{u_{ij}\}$, the fluxes are fully determined and the
Hamiltonian becomes quadratic in terms of the $c$ fermions.  The
ground state can be found by solving the quadratic Hamiltonians
corresponding to all possible flux configurations (or flux
\textit{sectors}) and identifying the flux sector that yields the
lowest energy state. Other flux sectors are important when considering the high-energy excitations and dynamic properties of the model\cite{Baskaran2007,Knolle2014}; however, we will limit our focus to the ground state properties of these models.

Unlike the 2D honeycomb lattice, both the \hhc{} (\textit{i.e.}
\hn{0}) and \hn{1} lattices possess loops without mirror symmetries.
As such, Lieb's theorem \cite{Lieb1994} cannot determine the flux
passing through these loops in the ground state.  We performed a
brute-force search throughout all flux sectors compatible with an
8-fold enlarged unit cell and the results suggest that the ground
state on the \hhc{} lattice belongs to the zero-flux sector, which
agrees with previous work\cite{Mandal2009}. In contrast, on the \hn{1}
lattice, we find that the ground state flux sector differs for
different values of $\delta=J_z/J_x$.  At the isotropic point
$\delta=1$, a particular flux configuration with $\pi$ flux passing
through a subset of the loops appears to be the ground state flux
sector (hereafter, we label it as the ``$\pi$-flux sector'').  Upon
increasing $\delta$, the zero-flux sector becomes energetically
favorable. We will first focus on the zero-flux sectors on the \hhc{}
and \hn{1} lattices and defer the more involved analysis of the
$\pi$-flux sector on the \hn{1} lattice for later.

\paragraph*{Bulk Majorana spectrum in the zero-flux sector: }Due to
the bipartite nature of both the \hhc{} and \hn{1} lattices, the
Hamiltonian in any flux sector takes the off-diagonal form
\begin{align}
  \label{eq:ham}
  \mathbf{H}^{\Phi}_n &= \sum_k {\vec{c}_{n,-k}}^{~T} H^{\Phi}_{n,k} \vec{c}_{n,k}\\
  H^{\Phi}_{n,k} &=
  \begin{bmatrix}
    0 & -iD^{\Phi}_{n,k} \\
    i\left({D^{\Phi}_{n,k}}\right)^\dag & 0
  \end{bmatrix},
\end{align}
where $n$ refers to the $n^{\text{th}}$-harmonic honeycomb, $\Phi$
labels the flux sector, and $\vec{c}_{n,k}$ is the vector of the
Fourier transforms of the $c$ Majorana fermions ordered by the odd
sublattices followed by the even sublattices (See Supplemental
Material \cite{supp} for definition of lattice vectors, unit cell, and
sublattice conventions).  In the zero-flux sector, we can choose the
gauge where $u_{ij} = 1$ when $i$ is an even sublattice and $j$ is an
odd sublattice. Consequently, the $D^{0}$-matrices for the \hhc{} and
\hn{1} lattices are
\begin{align}
  D^{0}_{0,k} =
  \begin{bmatrix}
    J_z & A_{k}e^{ik_3}\\
    B_{k} & J_z
  \end{bmatrix},
  D^{0}_{1,k} =
  \begin{bmatrix}
    J_z & 0 & 0 & A_{k}e^{ik_3} \\
    A_{k}^\ast & J_z & 0 & 0 \\
    0 & B_{k} & J_z & 0 \\
    0 & 0 & B_{k}^\ast & J_z
  \end{bmatrix},
\end{align}
where $A_{k} = J_x(1+e^{-ik_1})$, $B_{k} = J_x(1+e^{-ik_2})$ with $k_i=
\vec{k}\cdot \vec{a}_i$, and $\vec{a}_i$ are the lattice vectors.

\begin{figure*}
  \centering
  \setlength\fboxsep{0pt}
  \setlength\fboxrule{0pt}
  \begin{tabular}{ccccc}
    \multicolumn{2}{c}{
      \subfloat[\hn{0}: 0-flux, $\delta=1$]{
        \label{fig:hn_even}
        \fbox{\begin{overpic}[scale=1,clip=true,trim=0 -10 0 0]{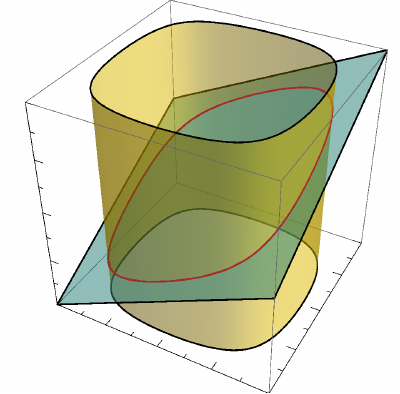}
            \put(30,8){$\scriptstyle{k_1}$}
            \put(104,25){$\scriptstyle{k_2}$}
            \put(0,102){$\scriptstyle{k_3}$}
            \put(42,16){$\scriptstyle{0}$}
            \put(12,26){$\scriptstyle{-\pi}$}
            \put(72,4){$\scriptstyle{\pi}$}
            \put(96,28){$\scriptstyle{0}$}
            \put(81,7){$\scriptstyle{-\pi}$}
            \put(106,48){$\scriptstyle{\pi}$}
            \put(5,61){$\scriptstyle{0}$}
            \put(3,36){$\scriptstyle{-\pi}$}
            \put(1,91){$\scriptstyle{\pi}$}
          \end{overpic}}
      }
    } &
    \subfloat[\hn{1}: 0-flux, $\delta=1$]{
      \label{fig:hn_odd}
      \fbox{\begin{overpic}[scale=1,clip=true,trim=0 -10 0 0]{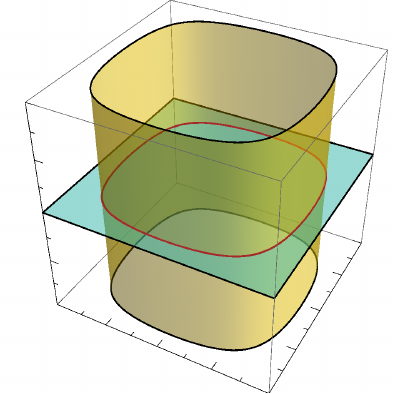}
          \put(30,8){$\scriptstyle{k_1}$}
          \put(104,25){$\scriptstyle{k_2}$}
          \put(0,102){$\scriptstyle{k_3}$}
          \put(42,16){$\scriptstyle{0}$}
          \put(12,26){$\scriptstyle{-\pi}$}
          \put(72,4){$\scriptstyle{\pi}$}
          \put(96,28){$\scriptstyle{0}$}
          \put(81,7){$\scriptstyle{-\pi}$}
          \put(106,48){$\scriptstyle{\pi}$}
          \put(5,61){$\scriptstyle{0}$}
          \put(3,36){$\scriptstyle{-\pi}$}
          \put(1,91){$\scriptstyle{\pi}$}
        \end{overpic}}
    } &
    \multicolumn{2}{c}{
      \subfloat[\hn{1}: $\pi$-flux, $\delta=\{1,7^{1/4}\}$]{
        \label{fig:h1_pi}
        \fbox{\begin{overpic}[scale=1,clip=true,trim=0 -10 0 0]{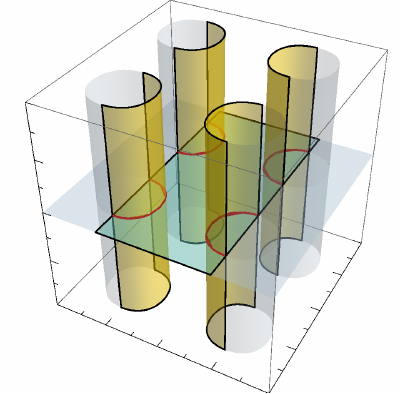}
            \put(30,8){$\scriptstyle{k_1}$}
            \put(104,25){$\scriptstyle{k_2}$}
            \put(0,102){$\scriptstyle{k_3}$}
            \put(42,16){$\scriptstyle{0}$}
            \put(12,26){$\scriptstyle{-\pi}$}
            \put(72,4){$\scriptstyle{\pi}$}
            \put(96,28){$\scriptstyle{0}$}
            \put(81,7){$\scriptstyle{-\pi}$}
            \put(106,48){$\scriptstyle{\pi}$}
            \put(5,61){$\scriptstyle{0}$}
            \put(3,36){$\scriptstyle{-\pi}$}
            \put(1,91){$\scriptstyle{\pi}$}
          \end{overpic}}
      }
    } \\
    \subfloat[(100) surface]{
      \label{fig:ss_h0_100}
      \fbox{\begin{overpic}[scale=1,clip=true,trim=-10 0 0 0,grid=false]{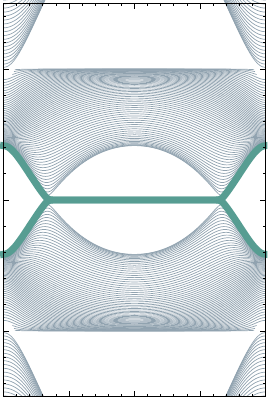}
          \put(5,75){\includegraphics[scale=1]{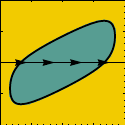}}
          \put(5,56){$\scriptstyle{0}$}
          \put(5,94){$\scriptstyle{1}$}
          \put(0,18){$\scriptstyle{-1}$}
          \put(0,120){$\scriptstyle{E}$ $\scriptstyle{(J_x)}$}
          \put(30,70){$\scriptstyle{k_2}$}
          \put(52,91){$\scriptstyle{k_3}$}
        \end{overpic}}
    } &
    \subfloat[(001) surface]{
      \label{fig:ss_h0_001}
      \fbox{\begin{overpic}[scale=1,clip=true,trim=0 0 0 0,grid=false]{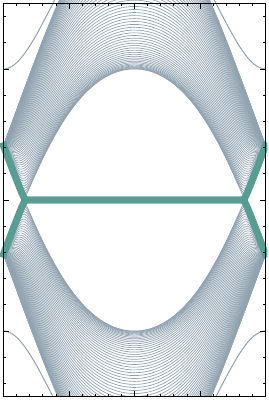}
          \put(5,75){\includegraphics[scale=1]{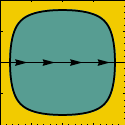}}
          \put(20,70){$\scriptstyle{k_1}$}
          \put(42,91){$\scriptstyle{k_2}$}
        \end{overpic}}
    } &
    \subfloat[(001) surface]{
      \label{fig:ss_h1_001}
      \fbox{\begin{overpic}[scale=1,clip=true,trim=-10 0 0 -15,grid=false]{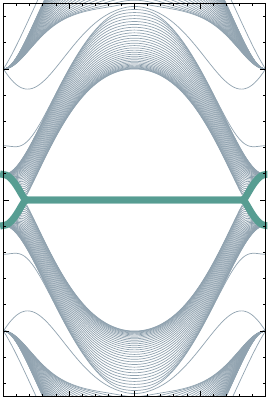}
          \put(5,75){\includegraphics[scale=1]{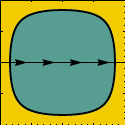}}
          \put(5,56){$\scriptstyle{0}$}
          \put(5,94){$\scriptstyle{1}$}
          \put(0,18){$\scriptstyle{-1}$}
          \put(0,120){$\scriptstyle{E}$ $\scriptstyle{(J_x)}$}
          \put(30,70){$\scriptstyle{k_1}$}
          \put(52,91){$\scriptstyle{k_2}$}
        \end{overpic}}
    } &
    \subfloat[(001) surface; $\delta=1$]{
      \label{fig:ss_h1_pi_001}
      \fbox{\begin{overpic}[scale=1,clip=true,trim=-10 0 0 0,grid=false]{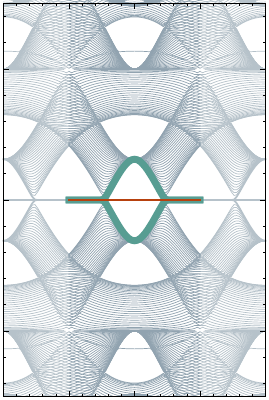}
          \put(5,75){\includegraphics[scale=1]{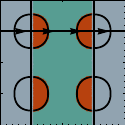}}
          \put(5,56){$\scriptstyle{0}$}
          \put(5,94){$\scriptstyle{1}$}
          \put(0,18){$\scriptstyle{-1}$}
          \put(30,70){$\scriptstyle{k_1}$}
          \put(52,91){$\scriptstyle{k_2}$}
          \put(0,18){$ $}
        \end{overpic}}
    } &
    \subfloat[(001) surface, $\delta=7^{1/4}$]{
      \label{fig:ss_h1_pi_001_2}
      \fbox{\begin{overpic}[scale=1,clip=true,trim=0 0 0 0,grid=false]{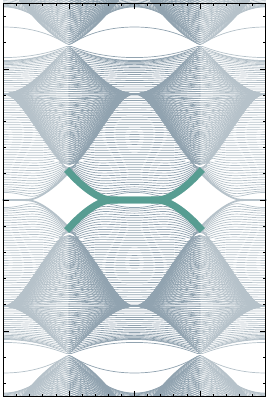}
          \put(5,75){\includegraphics[scale=1]{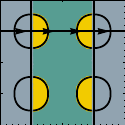}}
          \put(20,70){$\scriptstyle{k_1}$}
          \put(42,91){$\scriptstyle{k_2}$}
        \end{overpic}}
    }
    \end{tabular}
    \caption{(Color online) Position of nodal rings, surface spectra,
      and winding numbers.  In
      Figs. \ref{fig:hn_even}-\ref{fig:h1_pi}, the red lines show the
      location of the nodal rings of the indicated lattice, flux
      sector, and $\delta=J_z/J_x$.  The red lines are the
      intersections of the yellow and turquoise surfaces, which are
      the $\text{LHS}=1$ and $\text{RHS}=1$ of
      Eq. \ref{eq:hn_even_node}, Eq. \ref{eq:hn_odd_node}, and
      Eq. \ref{eq:h1_pi_node}.  In Fig. \ref{fig:h1_pi}, the first
      Brillouin zone spans the region $-\pi/2<k_1 \leq \pi/2$ due to
      doubling of the unit cell in the $\pi$-flux sector.
      Figs. \ref{fig:ss_h0_100}-\ref{fig:ss_h1_pi_001_2} show the
      surface spectra along one-dimensional momentum cuts on the
      various lattices and flux sectors, while the insets within
      indicate the location of the momentum cuts and projection of the
      nodal ring on the surfaces indicated.  The colors in the insets
      correspond to the winding numbers, where yellow, turquoise, and
      red are $\nu=0$, $\pm1$, and $\pm2$ respectively.  When
      $\nu\neq0$, as shown in
      Figs. \ref{fig:ss_h0_100}-\ref{fig:ss_h1_pi_001_2}, we find the
      presence of zero-energy surface flat bands with $|\nu|$-fold
      degeneracy, due to the bulk-boundary correspondence.}
\end{figure*}
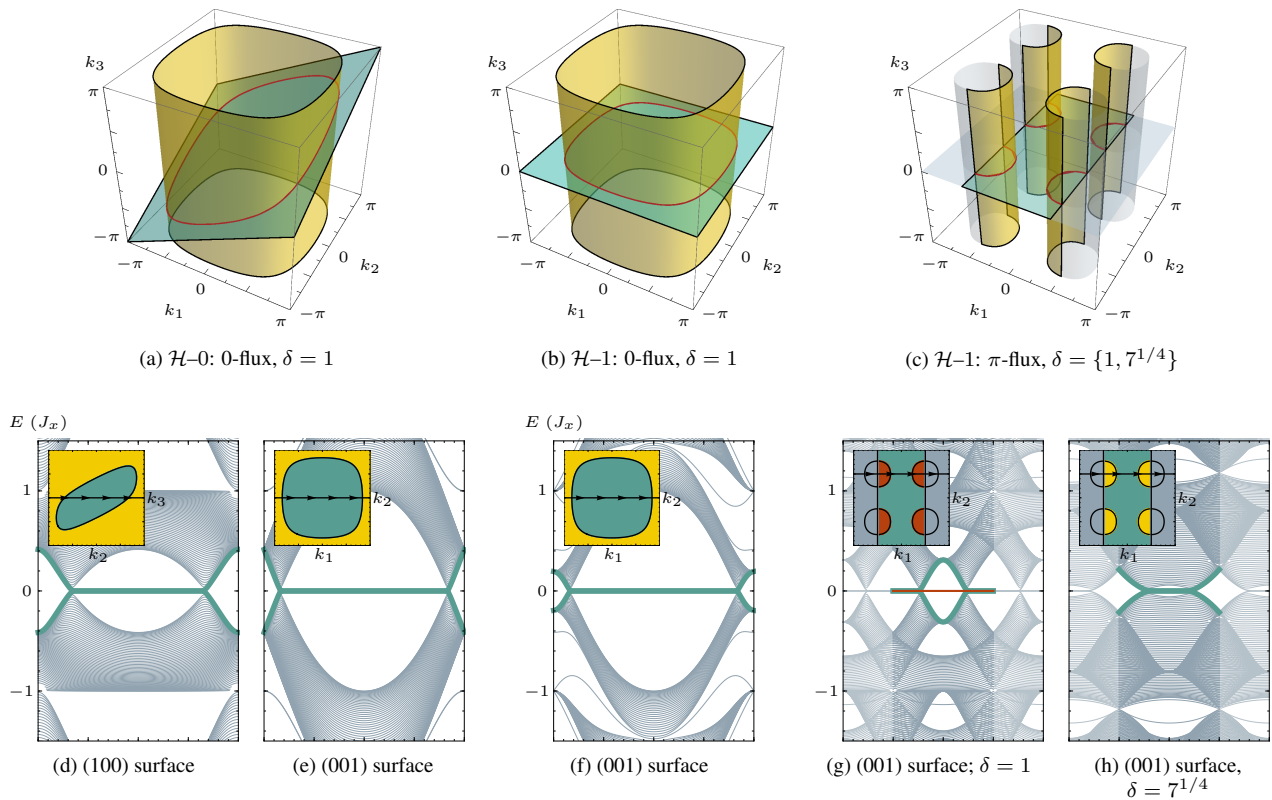

Each of the zero-flux sectors of both \hhc{} and \hn{1} lattices
possesses gapless spinon excitations in the bulk that form a nodal
ring in the 3D Brillouin zone (BZ).  The off-diagonal block form of
$H^{\Phi}_n$ ensures that the zero-modes of $H^{\Phi}_n$ are
determined by $\text{det}(D^{\Phi}_{n,k}) =0$.  For the zero-flux
phase of the \hhc{} and \hn{1} lattices, these conditions are
\begin{align}
  \label{eq:hn_even_node}
  &\text{\hn{0}}:\quad && \ \ 4\cos{\frac{k_1}{2}}\cos{\frac{k_2}{2}}\  = \delta^2 e^{-i(k_3 -
    \frac{k_1}{2} -
    \frac{k_2}{2})}, \\
  \label{eq:hn_odd_node}
  &\text{\hn{1}}: &&\left| 4\cos{\frac{k_1}{2}}\cos{\frac{k_2}{2}} \right| = \delta^2 e^{-i k_3}.
\end{align}
For values of $\delta<2$, a continuous set of solutions exist for each
of Eq. (\ref{eq:hn_even_node}) and Eq. (\ref{eq:hn_odd_node}), which
defines the nodal ring.  We have illustrated the locations of the
nodal rings for the isotropic case $\delta=1$ in
Fig. \ref{fig:hn_even} and Fig. \ref{fig:hn_odd}.

\paragraph*{Topological invariants of the nodal ring:} The nodal rings
present in the zero-flux sectors of the \hhc{} and \hn{1} models are
topologically stable.  To see this, we first define the time-reversal
(TR) and particle-hole (PH) symmetry operators, whose unitary
components satisfy the following relations
\begin{eqnarray}
  H_k=\epsilon_U U H^{T}_{-k} U^{-1},\quad~
  UU^{\dagger}=\mathbb{I},\quad~ U^{T}=\eta_U U,
\end{eqnarray}
where $H_k$ is the Hamiltonian matrix, $^T$ is the matrix transpose,
$\mathbb{I}$ is the identity matrix, $U=T,P$ for TR/PH respectively,
$\epsilon_{U}=\pm1$ for TR/PH, and $\eta_{U}=\pm1$.  The presence of
both TR and PH ensures that $\mathbf{S}=\mathbf{T}\mathbf{P}$ is a
chiral (or sublattice) symmetry of the system, which satisfies
$\{\mathbf{S},\mathbf{H}^{\Phi}_{n}\}=0$ (where boldface letters
denote operators).

In the case of the \hhc{} and \hn{1} lattices, we find for the
zero-flux sector
\begin{eqnarray}
  \label{eq:ops}
  T^{0}_{n} = S^{0}_{n} = \sigma_z \otimes \mathbb{I}_{2n+2}, \quad
  P^{0}_{n} = \mathbb{I}_{4n+4},
\end{eqnarray}
where $\sigma$ are the Pauli matrices, $\otimes$ is the tensor product
of matrices, and $\mathbb{I}_m$ is the $m\times m$ identity matrix.
In both systems, $\eta_T=\eta_P=+1$, which implies that $H^0_{n,k}$
belongs to symmetry class BDI based on the classification of
topologically stable Fermi surfaces (FS's)\cite{zhao2013topological,
  Matsuura2013}. The topological stability of a nodal ring in
three-dimensional systems of class BDI is characterized by the
following integer-valued topological invariant (winding number)
\begin{eqnarray}\label{eq:winding number}
  \nu = \frac{1}{4\pi i} \oint \text{d}k
  \text{Tr}[D_{\bk{}}^{-1}\partial_k D_{\bk{}} -
  (D^\dag)_{\bk{}}^{-1} \partial_k D_{\bk{}}^\dag ],
\end{eqnarray}
where the integral is taken along a path around the nodal ring.

We can deform the path into two pieces: one passing through the inside
of the nodal ring and one outside. Integrating Eq. \ref{eq:winding
  number} in the $k_3$ direction along the lines $k_1 = k_2 = 0$
(inside the nodal ring) and $k_1 = k_2 = \pi$ (outside the nodal
ring), we find a nontrivial winding number $\nu=1$ inside the nodal
ring but a trivial one ($\nu=0$) outside (See Supplemental Material
\cite{supp} for details). As a result, the nodal ring is characterized
by a topological index $\nu=\pm1$ and is hence topologically
stable. 

\paragraph{Surface spectra:} The surface spectra of the \hhc{} and \hn{1} lattices is expected to
possess zero-energy flat bands due to the bulk-boundary
correspondence\cite{Matsuura2013}, as long as the bulk nodal ring has
finite projection in the surface BZ. At the momenta corresponding to the projection of the nodal ring on a surface, the change in the number of flat bands must be the same as the topological index $\nu$ of the ring.

For the \hhc{} lattice, we examine the
spectra associated with the $(100)$ and $(001)$ surfaces in
Fig. \ref{fig:ss_h0_100} and \ref{fig:ss_h0_001} (the surface $(010)$
is related to the $(100)$ surface by a glide plane symmetry, hence it
is not shown).  Since the nodal ring has finite projection along $k_1$
and $k_3$, flat bands at zero energy are expected in both surface
spectra.  Indeed, we see $\nu=1$ within the area enclosed by the
projection of the nodal ring.  Plotting the surface spectra along
momentum paths that cut through the nodal ring projections, we see the
presence of flat bands where the winding number is $\pm1$.  In
contrast, the nodal ring in the \hn{1} lattice only has finite
projection along the $k_3$ direction.  Therefore, only the $(001)$
surface spectrum possesses zero energy flat bands, which can be seen in
Fig. \ref{fig:ss_h1_001}.

\paragraph*{Analysis of the $\pi$-flux sector:} The above analysis can
be performed analogously in the $\pi$-flux sector on the \hn{1}
lattice; here we summarize the main results.  The description of the
$\pi$-flux sector requires doubling of the unit cell in the $a_1$
direction (See Supplemental Material \cite{supp} for definition of the
enlarged unit cell and $D^{\pi}_{1,k}$).  Due to the enlarged unit
cell, the TR, PH, and chiral symmetry operators are now given by
\begin{eqnarray}
  \label{eq:ops_pi}
  T^{\pi}_{n} = S^{\pi}_{n} = \sigma_z \otimes \mathbb{I}_{4n+4}, \quad
  P^{\pi}_{n} = \mathbb{I}_{8n+8}
\end{eqnarray}
with $n=1$.  Since $\eta_{T}=\eta_{P}=+1$, $H^{\pi}_k$ still belongs
to class BDI and its nodal rings are associated with
$\mathbb{Z}$-valued topological invariants.

When $0<\delta<2^{3/4}$, the bulk spectrum
possesses two nodal rings that satisfy
\begin{eqnarray}
  \label{eq:h1_pi_node}
  16 \sin^2{k_1} \sin^2{k_2} = 8 e^{-i k_3}\delta^4 - e^{-2i k_3}\delta^8
\end{eqnarray}
where $-\frac{\pi}{2} \leq k_1 < \frac{\pi}{2}$ due to the doubling of
the unit cell.  The parameter point $\delta'\equiv\sqrt{2}$ is unique:
upon increasing $\delta$ towards this value, the two nodal rings
shrink towards $k'_{\pm}=(\frac{\pi}{2},\pm\frac{\pi}{2},0)$.  At
$\delta'$, the nodal rings turn into Dirac points at $k'_{\pm}$.  Upon
further increasing $\delta$ beyond $\delta'$, the nodal rings return
and expand.  For $\delta<\delta'$, $\nu=\pm2$ inside the nodal rings
and $\nu=\pm1$ outside the nodal rings.  On the other hand, when
$\delta'<\delta<2^{3/4}$, $\nu$ inside the nodal rings decreases to
$0$, while $\nu$ remains as $\pm1$ outside the nodal rings.  The
surface spectra of these cases are illustrated in
Fig. \ref{fig:ss_h1_pi_001} and \ref{fig:ss_h1_pi_001_2}.

\paragraph*{Generalization to the \hn{n} lattice:} The above results
for the zero-flux sector can be straightforwardly extended to all the
\hn{n} lattices\cite{Modic2014}.  In fact, many of the results remain the same: the
position of the nodal ring for even-$n$ lattices is given by
Eq. \ref{eq:hn_even_node}, while for the odd-$n$ lattices it's given
by Eq. \ref{eq:hn_odd_node}.  The operators $T^0_n$, $P^0_n$, and
$S^0_n$ are still defined by Eq. \ref{eq:ops} where $n$ is now
arbitrary, therefore the whole family of $H^0_{n,k}$ belongs to class
BDI.  For the calculation of the winding number, we have
\begin{eqnarray}
  \label{eq:trace}
  \text{Tr}[D_{n,k_3}^{-1}\partial_{k_3}D_{n,k_3}] =
  \frac{-ie^{ik_3}}{(\delta/2)^{2n+2}-e^{ik_3}}
\end{eqnarray}
along the line $k_1=k_2=0$ for all the \hn{n} lattices.  In addition,
$D^0_{n,k}$ is constant along $k_3$ when $k_1=k_2=\pi$.  Hence, the
winding number is always $1$ and 0 inside and outside the nodal ring,
respectively.

For the $\pi$-flux sector, the TR, PH, and chiral symmetry operators
are still defined by Eq. \ref{eq:ops_pi} for arbitrary $n$, and the
spinon Hamiltonian $H^{\pi}_{n,k}$ belongs to the BDI class.  Other
aspects of the Hamiltonian are less generalizable, however.  The zero
modes of the bulk Hamiltonian do not follow a generalized form;
however, we have numerically verified that two nodal rings are present
for $n<15$.  The point $\delta=\delta'$ remains as a special point
where the two nodal rings collapse to two points.  Like the \hn{1}
model, for $\delta<\delta'$, $\nu=2(1)$ inside (outside) the nodal
rings, while for $\delta>\delta'$ within the gapless phase, $\nu=0(1)$
inside (outside) the nodal rings. Therefore, the spinon nodal rings in
the bulk and surface flat bands are topologically protected.

\paragraph*{Summary and Discussion:} A nearest-neighbor tight-binding
Hamiltonian of spinless electrons hopping on a \hn{n} lattice will
have the same band structure as the zero-flux sector. With PH, chiral,
and charge conservation symmetry, it also belongs to symmetry class
BDI, with the unitary component of the time-reversal and particle-hole symmetry operators exchanged with respect to the Kitaev spin liquid. This means a half-filled electron system on the \hn{n} lattice
can also host topologically stable nodal rings in the bulk and
symmetry-protected flat bands on the surface.

Although this work focuses on the \hhc{} and \hn{n} lattices, our
analysis applies to Kitaev models on any bipartite and trivalent
lattice. Generally, the spinon band structure of any bipartite-lattice
Kitaev model belongs to symmetry class BDI, whose Hamiltonian has the
form of Eq. (\ref{eq:ham}) independent of the flux sector. Since the
Majorana spinon FS's are determined by two real equations
\begin{eqnarray}
  \label{eq:fermi surface constraint:BDI]}
  &\text{Re}\left[\det(D_k)\right]=\text{Im}\left[\det(D_k)\right]=0,
\end{eqnarray}
a $d$-dimensional lattice will generically give $(d-2)$-dimensional
spinon FS's\cite{Hermanns2014}. Similar to nodal rings in three-dimensional lattices,
each $(d-2)$-dimensional FS in the $d$-dimensional BZ is classified by
an integer-valued topological invariant $\nu$ for symmetry class
BDI. A non-zero $\nu$ will imply the stability of the spinon FS and
protected surface flat bands.

A simple example is the original Kitaev model on the honeycomb
lattice\cite{Kitaev2006}. Majorana spinons in the gapless
$\mathbb{Z}_2$ spin liquid ground state have a graphene-like band
structure, with a pair of topologically stable point nodes with
$\nu=\pm1$. This leads to localized spinon flat bands on the
boundaries.

An exception occurs when the surfaces defined by the two constraints
in Eq. (\ref{eq:fermi surface constraint:BDI]}) coincide, such as the
case of the hyperoctagon lattice\cite{Hermanns2014} where a 2D FS of
Majorana spinons arises. The FS is characterized by a $\mathbb{Z}_2$
topological index of class BDI and there are no surface flat bands
associated with it\cite{Matsuura2013}.

Near the isotropic limit of the \hn{n} models, we find the presence of
nodal rings. Through numerical and analytical calculations, we find
that the winding numbers around these rings are $\pm1$ in all cases
that we examined. As required by the bulk-boundary correspondence, we
find surface flat bands protected by the present symmetries.

In the strongly anisotropic limit, the nodal rings disappear and the
ground states of the \hn{n} models become gapped quantum spin liquids.
These gapped phases also have a nontrivial topology, characterized by
1D weak indices\cite{Ran2010} of symmetry class BDI. The physical
consequence of this weak index is that the surface flat bands of
Majorana spinons will persist even when we enter the gapped phase, as
long as translation symmetry is preserved\cite{Kou2008,Cho2012}. Once
we break TR and hence chiral symmetry (leading to symmetry class D),
the surface flat bands gain a dispersion and the bulk line node can gain a gap.

While this manuscript focuses on the properties of the model in the
ground state, we expect the results to extend to small finite
temperatures\cite{Nasu2014}. A gap exists to flux excitations in the
model, resulting in the number of such excitations being exponentially
supressed in the low temperature limit. As such, the band structure of
the spinon excitations is robust for small T, and the flat surface
band structure should be detectable experimentally in such a
system. Comparing the results of thermal transport measurements taken
across different surfaces allows one to identify the presence of such
surface modes, as these are absent on surfaces perpendicular to the
bulk nodal ring.
In addition, due to the divergence in the density of states on the surface at zero energy, we expect the surface contribution to the specific heat will dominate the $T^2$ bulk signal at sufficiently low temperatures\cite{Heikkila2011}.
By tuning the aspect ratio between the surface with flat bands and the other
surfaces (and the bulk), one may be able to isolate this contribution, providing strong evidence of the presence of these topological surface bands.


We thank Yige Chen for discussions. YML and YBK acknowledge the
hospitality of the Aspen Center for Physics (NSF Grant No. PHYS-
1066293), where some part of this work was performed.  This research
was supported by the NSERC, CIFAR and Centre for Quantum Materials at
the University of Toronto (RS,EKHL,YBK), and by Office of BES,
Materials Sciences Division of the U.S. DOE under contract
No. DE-AC02-05CH11231 (YML).

\pagebreak
\ \
\pagebreak
\widetext
\begin{center}
  \textbf{\large Supplemental Materials: Topological spinon semimetals
    and gapless boundary states in three dimensions}
\end{center}
\setcounter{equation}{0}
\setcounter{figure}{0}
\setcounter{table}{0}
\setcounter{page}{1}
\makeatletter
\renewcommand{\theequation}{S\arabic{equation}}
\renewcommand{\thefigure}{S\arabic{figure}}
\renewcommand{\bibnumfmt}[1]{[S#1]}
\renewcommand{\citenumfont}[1]{S#1}

\section{\label{app:lattice}Lattice}
We choose, for the \hn{n} lattice, the primitive lattice vectors
\begin{align}
\vec{a}_1 &= (1,-1,-2) \\
\vec{a}_2 &= (1,-1,2) \\
\vec{a}_3 &= \left\{ \begin{array}{ll}
(4,2,0)+(6,6,0)\times \frac{n}{2} &\qquad \text{if } n \text{ is even,}\\
(6,6,0)\times \frac{n+1}{2} & \qquad \text{if } n \text{ is odd.}
\end{array}\right.
\end{align}

For the sublattice positions, we choose the convention such that none
of the $z$-bonds connect sites between different unit cells.  In
particular, for the \hhc{} lattice, we define the sublattice positions as
\begin{align}
P_1 = (0,0,0), \ P_2 = (1,1,0), \  P_3 = (2,1,-1), \ P_4 = (3,2,-1),
\end{align}
while for the \hn{1} lattice, we define
\begin{align}
P_1 =& (0,0,0), \ \ \ \ P_2 = (1,1,0), \ \ \ \  P_3 = (1,2,-1), \  P_4 = (2,3,-1), \nonumber \\  P_5 =& (3,3,-2), \ P_6 = (4,4,-2), \
  P_7 = (4,5,-1),  \ P_8 = (5,6,-1).
\end{align}

\section{\label{app:solution}Computation of the winding number in the zero flux sector}

In the zero flux sector, the matrix $D^0_{n,k}$ takes the form
\begin{align}
D^0_{n,k} = \begin{bmatrix}
C_{11} & C_{12} \\
C_{21} & C_{22}
\end{bmatrix}, \ C_{11} = \begin{bmatrix}
J_z & 0 & 0 & \ldots \\
A_k^\ast & J_z & 0 & \ldots \\
0 & A_k & J_z & \ldots \\
\vdots & \vdots & \vdots & \ddots
\end{bmatrix}, \ C_{22} = \begin{bmatrix}
J_z & 0 & 0 & \ldots \\
B_k^\ast & J_z & 0 & \ldots \\
0 & B_k & J_z & \ldots \\
\vdots & \vdots & \vdots & \ddots
\end{bmatrix},
\end{align}
$C_{12}$ and $C_{21}$ are zero except for the top-right entry, which
takes the form $A_k e^{ik_3}$ and $B_k$ respectively, and the $C$'s
are $(n+1)\times (n+1)$ square matrices.

Along the line $k_1 = k_2 = 0$, $A_k = B_k = 2J_x$. The derivative
$\partial_{k_3}D^0_{n,k}$ is only non-zero in the top-right entry, on
which it has the value $2iJ_xe^{ik_3}$. As the quantity of interest is
the trace of the product of this matrix with the inverse of
$D^0_{n,k}$, we require the bottom-left component of the
inverse. Using the adjoint method, we can compute this as
\begin{align}
(D^0_{n,k})^{-1}_{2n+2,1} = \frac{\text{adj}(D^0_{n,k})_{2n+2,1}}{\det(D^0_{n,k})} = \frac{-(2J_x)^{2n+1}}{J_z^{2n+2} - (2J_x)^{2n+2}e^{ik_3}}
\end{align}
Combining these, we come to the result
\begin{align}
\text{Tr}[D_{n,k_3}^{-1}\partial_{k_3}D_{n,k_3}] =
  \frac{-ie^{ik_3}}{(\delta/2)^{2n+2}-e^{ik_3}}.
\end{align}
In order to integrate this from 0 to $2\pi$, we make the substitution
$x = e^{ik_3}$ and perform the contour integral. This evaluates to
$2\pi i$ if $\delta/2<1$ and 0 if $\delta/2>1$, corresponding to the
gapless and gapped phases of this model respectively.

The winding number, $\nu$, can be calculated using Eq.
(\ref{eq:trace}). As the second term is simply the complex conjugate
of the first, we find that $\nu = 1$ whenever the
ring is present, \textit{i.e.} for values of $\delta < 2$.

\section{\label{app:pi} Details of the $\pi$-flux phase}

In order to describe the $\pi$-flux sector, we must first double the
size of the unit cell, to a total of $8n+8$ sublattices. This can be
done in either the $\vec{a}_1$ or $\vec{a}_2$ direction; we choose the
$\vec{a}_1$ direction for concreteness. In this enlarged unit cell, we
label the sublattices from $1$ to $(4n+4)$ the same as those in the
zero-flux phase, and the sublattices from $(4n+5)$ to $(8n+8)$ as each
of the above translated by the vector $\vec{a}_1$.

In this expanded unit cell, the values of a subset of the $u_{ij}$
changes sign with respect to the zero flux sector, thus inserting
$\pi$ flux through certain loops in the lattice. The bonds on which
$u_{ij}$ changes sign are those that connect unit cells separated by
$\vec{a}_2$ and where $i,j$ are sublattices between $(4n+5)$ and
$(8n+8)$ (there are $n+1$ such bonds). On the \hn{1} lattice, the
result is that the loops whose fluxes are not constrained by Lieb's
theorem have $\pi$-flux passing through them.

Given the conventions chosen above, on the \hn{1} lattice the $D$
matrix for the $\pi$-flux phase takes the form
\begin{align}
D^\pi_{1,k} = \begin{bmatrix}
J_z & 0 & 0 & J_xe^{ik_3} & 0 & 0 & 0 & J_xe^{ik_3} \\
J_x & J_z & 0 & 0 & J_xe^{2ik_1} & 0 & 0 & 0 \\
0 & B_k^+ & J_z & 0 & 0 & 0 & 0 & 0 \\
0 & 0 & (B_k^+)^\ast & J_z & 0 & 0 & 0 & 0 \\
0 & 0 & 0 & J_xe^{i(k_3-2k_1)} & J_z & 0 & 0 & J_xe^{ik_3} \\
J_x & 0 & 0 & 0 & J_x & J_z & 0 & 0 \\
0 & 0 & 0 & 0 & 0 & B_k^- & J_z & 0 \\
0 & 0 & 0 & 0 & 0 & 0 & (B_k^-)^\ast & J_z
\end{bmatrix}
\end{align}
where $B_k^{\pm} = J_x(1\pm e^{-ik_2})$.
\end{document}